\documentclass[aps,preprint]{revtex4}%
\usepackage{amsfonts}
\usepackage{amsmath}
\usepackage{amssymb}
\usepackage{graphicx}
\usepackage{natbib}
\usepackage{chapterbib}
\providecommand{\U}[1]{\protect\rule{.1in}{.1in}}

\begin{document}
\bibliographystyle{unsrtnat}
\preprint{ }
\title{Classical Representation of a Quantum System at Equilibrium: Theory }
\author{James Dufty and Sandipan Dutta}
\affiliation{Department of Physics, University of Florida}

\begin{abstract}
A quantum system at equilibrium is represented by a corresponding
classical system, chosen to reproduce thermodynamic and structural
properties. The motivation is to allow application of classical
strong coupling theories and molecular dynamics simulation to
quantum systems at strong coupling. The correspondence is made at
the level of the grand canonical ensembles for the two systems. An
effective temperature, local chemical potential, and pair potential
are introduced to define the corresponding classical system. These
are determined formally by requiring the equivalence of the grand
potentials and their functional derivatives. Practical inversions of
these formal definitions are indicated via the integral equations
for densities and pair correlation functions of classical liquid
theory. Application to the ideal Fermi gas is demonstrated, and the
weak coupling form for the pair potential is given. In a companion
paper two applications are described: the thermodynamics and
structure of uniform jellium over a range of temperatures and
densities, and the shell structure of harmonically bound charges.

\end{abstract}
\date{01 Feb 2013}
\maketitle

\section{Introduction and Motivation}

\label{sec1}The equilibrium properties (thermodynamics and
structure) of classical systems for conditions of strong coupling
(e.g., liquids) can be described by a number of practical and
accurate approximations \cite{Hansen}. Examples are those based on
integral equations for the pair correlation function, such as the
Percus-Yevick and hypernetted chain approximations (HNC).
Thermodynamic properties are then determined from exact
representations in terms of the pair correlation function.
Extensions to inhomogeneous systems have been described in terms of
the pair correlation function as well. More extensive formulations
of the thermodynamics and structure for classical systems can be
obtained within the framework of classical density functional theory
\cite{Lutsko}. Molecular dynamic simulation (MD) is perhaps the most
accurate tool for the description of classical systems.

A corresponding practical description of strong coupling when
quantum effects (diffraction and degeneracy) are important is less
well-developed, except when a transformation to a corresponding weak
coupling quasi-particle representation can be found (e.g., phonons,
conventional Fermi liquid theory). Conditions of current interest
for which such representations are not available include those of
the growing class known as "warm, dense matter", i.e. complex
ion-electron systems \cite{WDM}. The ions are typically
semi-classical but the electron conditions span the full range of
classical plasmas to zero temperature solids. With such potential
applications in mind, the objective here is to extend the classical
approximation methods to include quantum effects. The idea is to
introduce a representative classical statistical mechanics that
embodies selected quantum effects via effective thermodynamic
parameters and effective pair potentials in the classical
Hamiltonian. These classical parameters and pair potential are
defined formally in such a way as to ensure the equivalence of the
classical and quantum thermodynamics and structure. In this way the
established effective classical methods can be applied to describe
quantum systems at strong coupling as well. The precise definition
of the effective classical statistical mechanics is given in the
next section. Of course, there is no possibility to map a quantum
system entirely onto an equivalent classical system. Instead, the
classical system is defined to yield only selective quantum
properties, in this case the thermodynamics and pair structure.
Other properties calculated from this effective statistical
mechanics, such as higher order structure or transport properties,
constitute uncontrolled approximations, and its utility in this more
general context must be obtained from experience with applications.

The analysis here is a formalization of related partial attempts to
incorporate quantum effects in classical calculations, most commonly
through effective classical pair potentials. There is an extensive
history for constructing such effective potentials
\cite{Uhlenbeck,Morita,Kelbg,Lado,Dunn67,Filinov,Dufty09}. Recent
reviews with references can be found in reference \cite{Jones07}.
Such potentials have been applied in a number of classical theories
and simulations. In particular they provide a definition for the
classical statistical mechanics of electron-ion systems which
otherwise do not exist due to the Coulomb collapse without
diffraction effects. Effective pair potentials defined from the two
particle density matrix do not assure that the thermodynamics or
structure will be given correctly by the corresponding classical
statistical mechanics, and those approaches provide uncontrolled
approximations in most applications. In contrast, the formal
approach here has the advantage of predicting the correct
thermodynamics and structure while retaining the simplicity of pair
potentials for point particles without approximation. However, the
additional complication implied by this is the need for effective
classical thermodynamic parameters (e.g., temperature and chemical
potential) that differ from those for the given quantum system.

The introduction of an effective temperature in this context appears
to have been done first by Perrot and Dharma-wardana (PDW) in their
classical map hypernetted chain model \cite{DWP00}. It provides a
means to calculate a pair correlation function $g_{c}(r)$ from the
classical HNC integral equation with two modifications: 1) the
actual pair potential is replaced by the sum of a pair potential to
give the correct ideal gas quantum pair correlation function, plus a
phenomenological modification of the actual pair interaction to
account for diffraction effects (Deutsch potential
\cite{Deutsch}), and 2) an effective temperature $T_{c}=\left(  T^{2}%
+T_{0}^{2}\right)  ^{1/2}$. The single new parameter $T_{0}$ is
determined as a function of the density by requiring that the
classical correlation energy at $T=0$ be the same as that for the
quantum system. The latter is determined from diffusion Monte Carlo
simulation data. The solution to the HNC integral equation with
these embedded quantum effects combined with classical strong
coupling provides an approximate $g_{c}(r)$ from which thermodynamic
properties can be calculated from their classical expressions (e.g.,
pressure from the virial equation). During the past decade this
remarkably simple and practical approach has been applied to a
number of quantum systems with surprising success for a wide range
of temperatures and densities. A recent review with references is
given in \cite{DW11}. These successes provide motivation and support
for the work presented here, which can be considered a formalization
of these concepts into a precisely defined context.

Most practical forms of present quantum many-body theories and
simulation methods have limited domains of validity, for example:
Fermi liquid theory (low temperature), quantum plasma methods (high
temperature), RPA with local field corrections (moderate coupling),
quantum density functional theory (low temperature), path integral
Monte Carlo (small particle number). As noted above these
limitations have become clear recently with attempts to describe new
experimental conditions of "warm, dense matter". The classical
formulation provides new access to these extreme conditions by
incorporating accurately described classical strong coupling effects
and hence provides a complementary new tool. In addition,
applications can often be numerically simpler (e.g., classical
density functional theory, molecular dynamics, classical Monte Carlo
simulation).

The definition of a representative classical system in terms of selected
equilibrium properties of the underlying quantum system is given in the next
section. The quantum system is represented in the grand canonical ensemble.
For generality an external single particle potential $\phi_{ext}\left(
\mathbf{r}\right)  $ is included so the thermodynamics is that for an
inhomogeneous system. Interactions among particles is given by a pair
potential $\phi\left(  \mathbf{r,r}^{\prime}\right)  $. Only a single
component system is considered here for simplicity. Thermodynamic properties
are functions of the temperature $T$ and functionals of the local chemical
potential $\mu\left(  \mathbf{r}\right)  =\mu-\phi_{ext}\left(  \mathbf{r}%
\right)  $. A corresponding classical system is specified in a
classical grand canonical ensemble in terms of an effective
classical temperature $T_{c}$, classical local chemical potential
$\mu_{c}\left(  \mathbf{r}\right)  =\mu _{c}-\phi_{c,ext}\left(
\mathbf{r}\right)  $, and classical pair potential $\phi_{c}\left(
\mathbf{r,r}^{\prime}\right)  $. These three classical quantities
are defined by three independent constraints equating chosen
classical and quantum thermodynamics and structure: equality of the
grand potentials and their functional derivatives with respect to
the chemical potential and pair potential for the classical and
quantum systems. Equivalently, this requires that the the pressures,
the local densities, and the pair correlation functions are the same
for the two systems.

Having defined the classical system in terms of equating classical
and quantum properties, there is then the practical problem of
inverting the defining equations for the classical temperature,
local chemical potential, and pair potential. Formally this is done
via classical density functional theory which relates the local
density and pair correlation function to the classical potential.
Inverting this and exploiting the equivalence of classical and
quantum pair correlation functions and densities gives the desired
expression for the classical pair potential as a functional of the
quantum pair correlation function and density. Examples of practical
implementations of this approach are given by Percus-Yevick (PY) and
HNC approximate integral equations. With the classical pair
potential determined (formally), the classical temperature is
obtained from the classical virial equation for the pressure and the
HNC equation for the chemical potential.

In this way the classical temperature, chemical potential, and pair
potential are given as explicit functionals of the quantum
properties. In practice, analysis of these functionals entails the
full quantum many-body problem so it would seem that little progress
has been made. The key assumption in this approach is that the
strong coupling effects have a dominant classical component while
dominant quantum effects are local (e.g. diffraction) or global but
weakly dependent on interactions (e.g. exchange degeneracy). Then,
simple weak coupling approximations to the defining functionals for
the classical parameters are sufficient while strong coupling
effects are obtained by a more complete implementation of the
classical statistical mechanics. A \ more ambitious approach would
try to include limited strong coupling effects in the classical
parameters (e.g. $T=0$ correlation energy as in the PDW model
above). The formalism here provides a basis for approximations that
can be tailored to the specific system or objective considered. A
brief overview of this approach has been described elsewhere \cite
{Dufty12}.

As a first application the uniform electron gas (jellium) is
considered in the following companion paper \cite{Dutta12}. The
effective pair potential used incorporates the exact ideal gas
exchange and the random phase weak coupling limit. The pair
correlation function is then obtained for this potential from the
HNC integral equation describing strong classical correlations,
known to be accurate for Coulomb interactions. The results are
compared to earlier quantum methods \cite{Tanaka86} and the above
PDW model over a range of temperatures and densities. Good agreement
with diffusion MC simulations at $T=0$ and recent restricted path
integral MC results at finite temperatures is obtained. A second
application described briefly there is to charges in a harmonic trap
where classical strong correlations produce shell structure.
Qualitative effects of diffraction and exchange on the formation of
shell structure are observed.

\section{Definition of the representative classical system}

\label{sec2}

\subsection{Thermodynamics from statistical mechanics}

Consider a quantum many-body system with Hamiltonian $H$ in a volume $V$ at
equilibrium described by the grand canonical ensemble, with inverse
temperature $\beta$ and chemical potential $\mu$ . For simplicity a one
component system is considered and dependence on internal degrees of freedom
such as spin is suppressed. The Hamiltonian $H$ is of the form%
\begin{equation}
H=K+\Phi+\Phi_{ext}, \label{2.1}%
\end{equation}
where $K$ denotes the total kinetic energy,
\begin{equation}
K=\sum_{i=1}^{N}\frac{p_{i}^{2}}{2m},\label{2.1a}%
\end{equation}
 $\Phi$ is the pair-wise
additive potential energy among particles, $\Phi_{ext}$ is an
external potential
coupling to each particle%
\begin{equation}
\Phi=\frac{1}{2}\sum_{ij}^{N}\phi(\mathbf{q}_{i},\mathbf{q}_{j}),\hspace
{0.2in}\Phi_{ext}=\sum_{i=1}^{N}\phi_{ext}\left(
\mathbf{q}_{i}\right)  ,
\label{2.2}%
\end{equation}
and $N$ is the particle number. The forms for the pair potential and
single particle external potential are left general for the present
(beyond well-known conditions for the existence of the grand
potential defined below). The statistical density operator for the
grand ensemble depends on $H$ in the form $H-\mu N$. It is
convenient to combine the contributions from the external potential
with that from $N$
\begin{equation}
H-\mu N=K+\Phi-\int d\mathbf{r}\mu(\mathbf{r})\widehat{n}(\mathbf{r}),
\label{2.4}%
\end{equation}
where $\widehat{n}(\mathbf{r})$ is the number density operator
\begin{equation}
\widehat{n}(\mathbf{r})=\sum_{i=1}^{N}\delta\left(  \mathbf{r}-\mathbf{q}%
_{i}\right)  , \label{2.5}%
\end{equation}
and $\mu(\mathbf{r})$ is a \emph{local} chemical potential
\begin{equation}
\mu(\mathbf{r})\equiv\mu-\phi_{ext}\left(  \mathbf{r}\right)  . \label{2.6}%
\end{equation}
A caret over $\widehat{n}$ is used to distinguish the operator from its
average value $n$.

The thermodynamics of the system is determined from the grand potential
$\Omega=-\beta^{-1}\ln\Xi$, where $\Xi$ is the grand partition function. The
thermodynamic parameters for this function are the inverse temperature $\beta
$, the local chemical potential $\mu(\mathbf{r})$, and the volume $V$. The
grand potential is
\begin{equation}
\Omega(\beta\mid\mu,\phi)=-\beta^{-1}\ln\Xi(\beta\mid\mu,\phi)=-\beta^{-1}%
\ln\sum_{N}Tr_{N}e^{-\beta\left(  K+\Phi-\int d\mathbf{r}\mu(\mathbf{r}%
)\widehat{n}(\mathbf{r})\right)  }. \label{2.7}%
\end{equation}
The symbol $Tr_{N}$ denotes a trace over the $N$ particle Hilbert
space with appropriate symmetry restrictions for Fermions or Bosons
(equivalently, in a second quantized representation the sum over N
and the $Tr_{N}$ represent a trace over Fock space).  The vertical
line in the arguments of $\Omega$ and $\Xi$ denotes a functional of
the quantity following it, e.g. $\Xi(\beta\mid\mu,\phi)$ is a
function of $\beta$ and a functional of $\mu(\mathbf{r})$ and
$\phi(\mathbf{q}_{i},\mathbf{q}_{j})$. Its functional dependence on
the pair potential $\phi(\mathbf{r},\mathbf{r}^{\prime})$ has also
been made explicit in this notation, although it is not strictly a
thermodynamic variable. The dependence on the volume $V$ is left
implicit. The
pressure $p(\beta\mid\mu)$ is proportional to the grand potential%
\begin{equation}
p(\beta\mid\mu,\phi)V=-\Omega(\beta\mid\mu,\phi). \label{2.8}%
\end{equation}

The first order derivatives of the grand potential give the internal energy
$E(\beta\mid\mu,\phi)$ and average number density $n(\mathbf{r;}\beta\mid\mu,\phi)$%
\begin{equation}
E(\beta\mid\mu,\phi)=\frac{\partial\beta\Omega(\beta\mid\mu,\phi)}%
{\partial\beta}\mid_{\beta\mu(\mathbf{r})}, \label{2.9}%
\end{equation}%
\begin{equation}
n(\mathbf{r;}\beta\mid\mu,\phi)V=-\frac{\delta\Omega(\beta\mid\mu,\phi
)}{\delta\mu(\mathbf{r})}\mid_{\beta}. \label{2.10}%
\end{equation}
Higher order derivatives provide the fluctuations (susceptibilities)
and structure. In particular, the second functional derivative with
respect to $\mu(\mathbf{r})$ is related to the response
function $\chi(\mathbf{r},\mathbf{r}^{\prime}\mathbf{;}\beta\mid\mu)$%
\begin{align}
\frac{1}{\beta}\frac{\delta^{2}\Omega(\beta\mid\mu,\phi)}{\delta
\mu(\mathbf{r})\delta\mu(\mathbf{r}^{\prime})}  &  \mid_{\beta}%
=-\chi(\mathbf{r},\mathbf{r}^{\prime}\mathbf{;}\beta\mid\mu,\phi)\nonumber\\
&  =-\frac{1}{\beta}\int_{0}^{\beta}d\beta^{\prime}\left\langle e^{\beta
^{\prime}H}\delta\widehat{n}(\mathbf{r})e^{-\beta^{\prime}H}\widehat
{n}(\mathbf{r}^{\prime});\beta\mid\mu,\phi\right\rangle , \label{2.10a}%
\end{align}
where $\delta \widehat {n}(\mathbf{r})=\widehat
{n}(\mathbf{r})-n(\mathbf{r})$, and $\left\langle
X;\beta\mid\mu,\phi\right\rangle $ denotes an equilibrium grand
canonical average of the quantity $X$. A related quantity is the
pair correlation function,
$g(\mathbf{r},\mathbf{r}^{\prime}\mathbf{;}\beta\mid \mu,\phi)$,
obtained by functional differentiation with respect to the pair
potential%
\begin{align}
\frac{\delta\Omega(\beta\mid\mu,\phi)}{\delta\phi
(\mathbf{r},\mathbf{r}^{\prime})}\mid_{\beta,\mu} &\equiv
n(\mathbf{r;}\beta\mid\mu,\phi)
n(\mathbf{r}^{\prime}\mathbf{;}\beta\mid
\mu,\phi)g(\mathbf{r},\mathbf{r}^{\prime}\mathbf{;}\beta\mid\mu
,\phi)\nonumber\\
&=\left\langle\widehat{n}(\mathbf{r})\widehat
{n}(\mathbf{r}^{\prime});\beta\mid\mu,\phi\right\rangle -n(\mathbf{r;}%
\beta\mid\mu,\phi)\delta\left(
\mathbf{r}-\mathbf{r}^{\prime}\right)
\label{2.10c}%
\end{align}
For the quantum system there is no simple relationship between the two
measures of structure, $\chi(\mathbf{r},\mathbf{r}^{\prime}\mathbf{;}\beta
\mid\mu,\phi)$ and $g(\mathbf{r},\mathbf{r}^{\prime}\mathbf{;}\beta\mid
\mu,\phi)$. Instead, only their time dependent extensions are related via a
fluctuation - dissipation relation.

A corresponding classical system is considered with Hamiltonian $H_{c}$ in the
same volume $V$ at equilibrium described by the classical grand canonical
ensemble, with inverse temperature $\beta_{c}$ and local chemical potential
$\mu_{c}(\mathbf{r})$ . The Hamiltonian has the same form as (\ref{2.1})
except that the potential energy functions (\ref{2.2}) are
different, and denoted by%
\begin{equation}
\Phi_{c}=\frac{1}{2}\sum_{i\neq j}^{N}\phi_{c}(\mathbf{q}_{i},\mathbf{q}%
_{j}),\hspace{0.2in}\Phi_{c,ext}=\sum_{i=1}^{N}\phi_{c,ext}\left(
\mathbf{q}_{i}\right)  . \label{2.11}%
\end{equation}
The local chemical potential is $\mu_{c}(\mathbf{r})$%
\begin{equation}
\mu_{c}(\mathbf{r})\equiv\mu_{c}-\phi_{c,ext}\left(  \mathbf{r}\right)  .
\label{2.13}%
\end{equation}
The classical grand potential is defined in terms of these quantities by%
\begin{equation}
\beta\Omega_{c}(\beta_{c}\mid\mu_{c},\phi_{c})=-\ln\Xi_{c}(\beta_{c}\mid
\mu_{c},\phi_{c})=-\ln\sum_{N}\frac{1}{\lambda_{c}^{3N}N!}\int d\mathbf{q}%
_{1}..d\mathbf{q}_{N}e^{-\beta_{c}\left(  \Phi_{c}-\int dr\mu_{c}%
(r)\widehat{n}(r)\right)  }. \label{2.14}%
\end{equation}
Here, $\lambda_{c}=\left(  2\pi\beta_{c}\hbar^{2}/m\right)  ^{1/2}$ is the
thermal de Broglie wavelength associated with the classical temperature. The
integration for the partition function is taken over the $N$ particle
configuration space.

The classical thermodynamics is determined in the same way as in (\ref{2.8})
- (\ref{2.10})%
\begin{equation}
p_{c}(\beta_{c}\mid\mu_{c},\phi_{c})V=-\Omega_{c}(\beta_{c}\mid\mu_{c}%
,\phi_{c}). \label{2.15}%
\end{equation}%
\begin{equation}
E_{c}(\beta_{c}\mid\mu_{c},\phi_{c})=\frac{\partial\beta_{c}\Omega_{c}%
(\beta_{c}\mid\mu_{c},\phi_{c})}{\partial\beta_{c}}\mid_{\beta_{c}\mu_{c},\phi_{c}%
}, \label{2.16}%
\end{equation}%
\begin{equation}
n_{c}(\mathbf{r;}\beta_{c}\mid\mu_{c},\phi_{c})=-\frac{\delta\Omega(\beta
_{c}\mid\mu_{c},\phi_{c})}{\delta\mu_{c}(\mathbf{r})}\mid_{\beta_{c},\phi_{c}}. \label{2.17}%
\end{equation}%
\begin{align}
\frac{1}
{\beta_{c}}\frac{\delta^{2}\Omega(\beta_{c}\mid\mu_{c},\phi_{c})}{\delta
\mu_{c}(\mathbf{r})\delta\mu_{c}(\mathbf{r}^{\prime})}\mid_{\beta_{c},\phi
_{c}}& \equiv
-\chi_{c}(\mathbf{r},\mathbf{r}^{\prime}\mathbf{;}\beta_{c}\mid\mu_{c},\phi
_{c}) \nonumber\\
&=-\left\langle \delta\widehat{n}(\mathbf{r})\widehat{n}(\mathbf{r}
^{\prime});\beta_{c}\mid\mu_{c},\phi_{c}\right\rangle _{c} \label{2.17a}%
\end{align}%
\begin{align}
\frac{\delta\Omega(\beta_{c}\mid\mu_{c},\phi_{c})}{\delta
\phi_{c}(\mathbf{r},\mathbf{r}^{\prime})}\mid_{\beta_{c},\mu_{c}}&\equiv
n_{c}(\mathbf{r;}\beta_{c}\mid\mu_{c},\phi_{c})n_{c}(\mathbf{r}^{\prime
}\mathbf{;}\beta_{c}\mid\mu_{c},\phi_{c}) g_{c}(\mathbf{r}
,\mathbf{r}^{\prime}\mathbf{;}\beta_{c}\mid\mu_{c},\phi_{c}) \nonumber  \\
&=\left\langle \delta\widehat{n}(\mathbf{r})\widehat{n}(\mathbf{r}
^{\prime});\beta_{c}\mid\mu_{c},\phi_{c}\right\rangle _{c}-n_{c}
(\mathbf{r;}\beta_{c}\mid\mu_{c},\phi_{c})\delta\left(  \mathbf{r}
-\mathbf{r}^{\prime}\right).
\label{2.17b}%
\end{align}

Here $\left\langle X;\beta_{c}\mid\mu_{c},\phi_{c}\right\rangle _{c}$ denotes
the corresponding classical equilibrium grand canonical average of the
quantity $X$. Note the additional constraint that the derivatives are taken at
constant pair potential $\phi_{c}$. This is necessary because the
determination of $\beta_{c}$, $\mu_{c}(\mathbf{r})$, and $\phi_{c}$ for the
quantum correspondence in the next subsection implies $\phi_{c}(\mathbf{r}%
,\mathbf{r}^{\prime})$ is a function of $\beta_{c}$ and a functional
of $\mu_{c}(\mathbf{r})$. This dependence is left implicit to
simplify the notation. See Section \label{sec3} below for further
elaboration.

\subsection{Classical - quantum correspondence conditions}

The classical system has undefined ingredients: the effective inverse
temperature, $\beta_{c}$, the local chemical potential, $\mu_{c}(\mathbf{r})$,
and the pair potential for interaction among the particles, $\phi
_{c}(\mathbf{r},\mathbf{r}^{\prime})$. A correspondence between the classical
and quantum systems is defined by expressing these quantities as functions or
functionals of $\beta$, $\mu(\mathbf{r})$, and $\phi(\mathbf{r},\mathbf{r}%
^{\prime})$. This is accomplished by requiring the numerical equivalence of
two independent thermodynamic properties and one structural property for the
classical and quantum systems. The first two are chosen to be the equivalence
of the grand potential and its first functional derivative with respect to the
local chemical potential.
\begin{equation}
\Omega_{c}(\beta_{c}\mid\mu_{c},\phi_{c})\equiv\Omega(\beta\mid\mu
,\phi),\hspace{0.25in}\frac{\delta\Omega_{c}(\beta_{c}\mid\mu_{c},\phi_{c}%
)}{\delta\mu_{c}(\mathbf{r})}\mid_{\beta_{c},\phi_{c}}\equiv\frac{\delta
\Omega(\beta\mid\mu,\phi)}{\delta\mu(\mathbf{r})}\mid_{\beta}. \label{2.18}%
\end{equation}
An equivalent form for these conditions can be given in terms of the pressure
and density%
\begin{equation}
p_{c}(\beta_{c}\mid\mu_{c},\phi_{c})\equiv p(\beta\mid\mu,\phi),\hspace
{0.25in}n_{c}(\mathbf{r;}\beta_{c}\mid\mu_{c},\phi_{c})\equiv n(\mathbf{r;}%
\beta\mid\mu,\phi). \label{2.19}%
\end{equation}
These two relations provide two independent relations between of $\beta_{c}$,
$\mu_{c}(\mathbf{r})$ and the physical variables $\beta$ and $\mu(\mathbf{r})$.

It remains to have a structural equivalence to relate the pair potential
$\phi_{c}(\mathbf{r},\mathbf{r}^{\prime})$ to $\phi(\mathbf{r},\mathbf{r}%
^{\prime})$, which are two particle functions. This is accomplished by
equating the functional derivatives of the grand potentials with respect to
these pair functions
\begin{equation}
\frac{\delta\Omega_{c}(\beta_{c}\mid\mu_{c},\phi_{c})}{\delta\phi
_{c}(\mathbf{r},\mathbf{r}^{\prime})}\mid_{\beta_{c},\mu_{c}}=\frac
{\delta\Omega(\beta\mid\mu,\phi)}{\delta\phi(\mathbf{r},\mathbf{r}^{\prime}%
)}\mid_{\beta,\mu}. \label{2.20}%
\end{equation}
More physically, this implies the equivalence of density fluctuations%
\begin{equation}
\left\langle \delta\widehat{n}(\mathbf{r})\widehat{n}(\mathbf{r}^{\prime
});\beta_{c}\mid\mu_{c},\phi_{c}\right\rangle _{c}\equiv\left\langle
\delta\widehat{n}(\mathbf{r})\widehat{n}(\mathbf{r}^{\prime});\beta\mid
\mu,\phi\right\rangle . \label{2.20a}%
\end{equation}
Finally, from the equivalence of densities and the definitions of pair
correlation functions in (\ref{2.10c}) and (\ref{2.17b}) this third condition
becomes%
\begin{equation}
g_{c}(\mathbf{r},\mathbf{r}^{\prime};\beta_{c}\mid\mu_{c},\phi_{c})\equiv
g(\mathbf{r},\mathbf{r}^{\prime};\beta\mid\mu,\phi). \label{2.20b}%
\end{equation}
In summary, the classical - quantum correspondence conditions are the
equivalence of the pressures, densities, and pair correlation functions.

It might seem that a natural alternative choice to (\ref{2.20b})
would be to equate the second functional derivatives of the grand
potential with respect
to the local chemical potential, or equivalently $\chi_{c}(\mathbf{r}%
,\mathbf{r}^{\prime}\mathbf{;}\beta_{c}\mid\mu_{c},\phi_{c})\equiv
\chi(\mathbf{r},\mathbf{r}^{\prime}\mathbf{;}\beta\mid\mu,\phi)$. However, the
classical response function has a singular contribution proportional to
$\delta\left(  \mathbf{r}-\mathbf{r}^{\prime}\right)  $ that is not present in
the quantum response function. Hence their equivalence does not provide a
simple mapping of the parameters. In contrast, the classical and quantum forms
for the pair correlation functions are similar and do not have this problem.

In this way the three equations of (\ref{2.19}) and (\ref{2.20b}) determine,
formally, the classical parameters $\beta_{c},$ $\mu_{c},$ and $\phi
_{c}\left(  q\right)  $ as functions of $\beta,$ and functionals of
$\mu(\mathbf{r})$, and $\phi\left( \mathbf{r,r}^{\prime}\right)  $ $\hspace{0.25in}$%
\begin{equation}
\beta_{c}=\beta_{c}(\beta\mid\mu,\phi),\hspace{0.25in}\mu_{c}=\mu_{c}%
(\mathbf{r;}\beta\mid\mu,\phi),\hspace{0.25in}\phi_{c}=\phi_{c}\left(
\mathbf{r,r}^{\prime};\beta\mid\mu,\phi\right)  . \label{2.21}%
\end{equation}
This completes the definition of the classical system representative of the
given quantum system.

\subsection{Inversion of the correspondence conditions}

To determine the classical parameters, (\ref{2.21}), inversion of the
functional forms (\ref{2.19}) and (\ref{2.20b}) is required. This is done on
the basis of three exact results of classical statistical mechanics, for the
pressure, density, and pair correlation function respectively. The first is
the virial form for the classical pressure (obtained by differentiating
$\Omega_{c}(\beta_{c}\mid\mu_{c},\phi_{c})$ with respect to the volume)%
\begin{equation}
\beta_{c}p_{c}(\beta_{c}\mid\mu_{c},\phi_{c})=\frac{1}{V}\int d\mathbf{r}%
n_{c}(\mathbf{r})\left[
1+\frac{1}{3}\mathbf{r}\cdot\nabla\beta_{c}\mu
_{c}\left(  \mathbf{r}\right)  -\frac{1}{6}\int d\mathbf{r}^{\prime}%
n_{c}(\mathbf{r}^{\prime})g_{c}(\mathbf{r,r}^{\prime})\mathbf{r}^{\prime}%
\cdot\nabla^{\prime}\beta_{c}\phi_{c}(\mathbf{r,r}^{\prime})\right]
\label{2.22}%
\end{equation}
(Note that this is the average pressure over the entire system. A local
pressure could be identified with the integrand of (\ref{2.22})). The second
is an equation for the classical densities,%
\begin{equation}
\ln\left(  n_{c}\left(  \mathbf{r}\right)  \lambda_{c}^{3}\right)  =\beta
_{c}\mu_{c}(\mathbf{r})+\int_{0}^{1}d\alpha\int d\mathbf{r}^{\prime\prime
}c_{c}^{(2)}(\mathbf{r,r}^{\prime\prime}\mid\alpha n_{c})n_{c}\left(
\mathbf{r}^{\prime\prime}\right)  . \label{2.23}%
\end{equation}
and the third is an equation for the pair correlation functions%
\begin{align}
\ln g_{c}(\mathbf{r,r}^{\prime})  &  =-\beta_{c}\phi_{c}(\mathbf{r}%
,\mathbf{r}^{\prime})+\int_{0}^{1}d\alpha\int d\mathbf{r}^{\prime\prime}%
c_{c}^{(2)}(\mathbf{r,r}^{\prime\prime}\mid n_{c}+\alpha n_{c}\left(
g_{c}-1\right)  )\nonumber\\
&  \times n_{c}\left(  \mathbf{r}^{\prime\prime}\right)  \left(
g_{c}(\mathbf{r}^{\prime\prime}\mathbf{,r}^{\prime})-1\right)  \label{2.24}%
\end{align}
The direct correlation function $c_{c}^{(2)}$ appearing in these
equations
is determined from $g_{c}$ via the Ornstein-Zernicke equations%
\begin{align}
\left(  g_{c}\left(  \mathbf{r,r}^{\prime}\right)  -1\right)   &
=c_{c}\left(  \mathbf{r,r}^{\prime}\mid n_{c}\right) \nonumber\\
&  +\int d\mathbf{r}^{\prime\prime}c_{c}\left(  \mathbf{r,r}^{\prime\prime
}\mid n_{c}\right)  n_{c}\left(  \mathbf{r}^{\prime\prime}\right)  \left(
g_{c}\left(  \mathbf{r}^{\prime\prime}\mathbf{,r}^{\prime}\right)  -1\right)
. \label{2.25}%
\end{align}
The constant $\lambda_{c}$ in (\ref{2.23}) is the thermal de Broglie
wavelength evaluated at the classical temperature,
$\lambda_{c}=\left( 2\pi{\hbar}^{2}\beta_{c}/m\right) ^{1/2}$. See
Appendix \ref{apA} for further elaboration.

These equations provide the inversion in the following way. First,
(\ref{2.23}) and (\ref{2.24}) are solved for $\beta_{c}\mu_{c}\left(
\mathbf{r}\right)  $ and $\beta_{c}\phi_{c}(\mathbf{r},\mathbf{r}^{\prime})$
as functionals of $n_{c}(\mathbf{r})$ and $g_{c}(\mathbf{r},\mathbf{r}%
^{\prime})$. Then $n_{c}(\mathbf{r})$ and $g_{c}(\mathbf{r},\mathbf{r}%
^{\prime})$ are replaced by their quantum counterparts, $n(\mathbf{r})$ and
$g(\mathbf{r},\mathbf{r}^{\prime})$, according to the definitions (\ref{2.19})
and (\ref{2.20b}), giving the desired exact expressions%
\begin{equation}
\beta_{c}\mu_{c}(\mathbf{r})=\frac{3}{2}\ln\left(  \frac{\beta_{c}}{\beta
}\right)  +\ln\left(  n\left(  \mathbf{r}\right)  \lambda^{3}\right)
-\int_{0}^{1}d\alpha\int d\mathbf{r}^{\prime\prime}c^{(2)}(\mathbf{r,r}%
^{\prime\prime}\mid\alpha n)n\left(  \mathbf{r}^{\prime\prime}\right)  .
\label{2.26}%
\end{equation}%
\begin{align}
\beta_{c}\phi_{c}(\mathbf{r},\mathbf{r}^{\prime})  &  =-\ln g(\mathbf{r,r}%
^{\prime})+\int_{0}^{1}d\alpha\int d\mathbf{r}^{\prime\prime}c^{(2)}%
(\mathbf{r,r}^{\prime\prime}\mid n+\alpha n\left(  g-1\right)  )\nonumber\\
&  \times n\left(  \mathbf{r}^{\prime\prime}\right)  \left(  g(\mathbf{r}%
^{\prime\prime}\mathbf{,r}^{\prime})-1\right)  . \label{2.27}%
\end{align}
Here, $c^{(2)}(\mathbf{r,r}^{\prime\prime}\mid n)$ is defined by the
Ornstein-Zernicke equation (\ref{2.25}) with $n_{c}\left(  \mathbf{r}\right)
,g_{c}(\mathbf{r,r}^{\prime})$ replaced by $n\left(  \mathbf{r}\right)
,g(\mathbf{r,r}^{\prime})$%
\begin{equation}
c^{(2)}\left(  \mathbf{r,r}^{\prime}\mid n\right)  =\left(  g\left(
\mathbf{r,r}^{\prime}\right)  -1\right)  -\int
d\mathbf{r}^{\prime\prime }c^{(2)}\left(
\mathbf{r,r}^{\prime\prime}\mid n\right)  n\left(
\mathbf{r}^{\prime\prime}\right)  \left( g\left(
\mathbf{r}^{\prime\prime
}\mathbf{,r}^{\prime}\right)  -1\right)  . \label{2.27a}%
\end{equation}
Finally, an equation for $\beta_{c}/\beta$ is obtained by using the
equivalence of the classical and quantum pressures $p_{c}=p$,
(\ref{2.19}),
to write%
\begin{equation}
\frac{\beta_{c}}{\beta}=\frac{\beta_{c}p_{c}}{\beta p}. \label{2.28}%
\end{equation}
or with (\ref{2.22})%
\begin{equation}
\frac{\beta_{c}}{\beta}=\frac{1}{\beta pV}\int
d\mathbf{r}n(\mathbf{r})\left[
1+\frac{1}{3}\mathbf{r}\cdot\nabla\beta_{c}\mu_{c}\left(
\mathbf{r}\right)
-\frac{1}{6}\int d\mathbf{r}^{\prime}n(\mathbf{r}^{\prime})g(\mathbf{r,r}%
^{\prime})\mathbf{r}^{\prime}\cdot\nabla^{\prime}\beta_{c}\phi_{c}(\left\vert
\mathbf{r-r}^{\prime}\right\vert )\right]  . \label{2.29}%
\end{equation}

Equations (\ref{2.26}), (\ref{2.27}), (\ref{2.27a}), and (\ref{2.29}) then
determine the classical parameters as $\beta_{c}/\beta$, $\beta_{c}\mu
_{c}(\mathbf{r})$, and $\beta_{c}\phi_{c}(\mathbf{r},\mathbf{r}^{\prime})$
from the properties of the given quantum system%
\begin{equation}
\beta_{c}=\beta_{c}\left(  \beta\mid n,g\right)  ,\hspace{0.25in}\mu
_{c}(\mathbf{r})=\mu_{c}(\mathbf{r},\beta\mid n,g),\hspace{0.25in}\phi
_{c}(\mathbf{r,r}^{\prime})=\phi_{c}(\mathbf{r,r}^{\prime},\beta\mid n,g)
\label{2.30}%
\end{equation}
which is equivalent to (\ref{2.21}) (corresponding to a change of variables
from $\mu,\phi$ to $n,g$ for the quantum system).

These expressions for the classical parameters are exact, but deceptively
simple, equations. Their complexity arises from the fact that different
arguments for the functional $c^{(2)}(\mathbf{r,r}^{\prime\prime}\mid\cdot)$
occur. The relationship of two functionals at different arguments depends on
higher order correlations that are not determined by these equations. The
many-body problem has not been solved, only hidden, and approximations are
required for practical applications

The application of these results proceeds in three steps. First an
approximation to the classical equations (\ref{2.23}) and (\ref{2.24}) is
chosen such that classical correlations are well described. These are then
inverted to obtain the corresponding approximations to (\ref{2.26}) and
(\ref{2.27}). Next, selected information about the quantum system is chosen in
the form of approximations to $p,n,$ and $g$, for the calculation of the
classical parameters of (\ref{2.30}). Finally, with all classical parameters
known properties of interest are determined from classical statistical
mechanics (e.g., liquid state integral equations, classical density functional
theory, molecular dynamics).

\subsection{Hypernetted Chain Approximation}

To illustrate the first step, selection of a classical approximation for
(\ref{2.23}) and (\ref{2.24}), the HNC is noted. This results from making them
local in the function space of densities%
\begin{equation}
c_{c}^{(2)}(\mathbf{r,r}^{\prime\prime}\mid\alpha n_{c})\rightarrow
c_{c}^{(2)}(\mathbf{r,r}^{\prime\prime}\mid n_{c}),\hspace{0.2in}c_{c}%
^{(2)}(\mathbf{r,r}^{\prime\prime}\mid n_{c}+\alpha n_{c}\left(
g_{c}-1\right)  )\rightarrow c_{c}^{(2)}(\mathbf{r,r}^{\prime\prime}\mid
n_{c}). \label{2.31}%
\end{equation}
Then equations (\ref{2.23}) and (\ref{2.24}) become%
\begin{equation}
\ln\left(  n_{c}\left(  \mathbf{r}\right)  \lambda_{c}^{3}\right)  =\beta
_{c}\mu_{c}(\mathbf{r})+\int d\mathbf{r}^{\prime}c_{c}^{(2)}(\mathbf{r,r}%
^{\prime}\mid n_{c})n_{c}\left(  \mathbf{r}^{\prime\prime}\right)  .
\label{2.32}%
\end{equation}
\begin{equation}
\ln g_{c}(\mathbf{r,r}^{\prime}\mid n_{c})=-\beta_{c}\phi_{c}(\mathbf{r}%
,\mathbf{r}^{\prime})+\int d\mathbf{r}^{\prime\prime}c_{c}^{(2)}%
(\mathbf{r,r}^{\prime\prime}\mid n_{c})n_{c}\left(  \mathbf{r}^{\prime\prime
}\right)  \left(  g_{c}(\mathbf{r}^{\prime\prime}\mathbf{,r}^{\prime}\mid
n_{c})-1\right)  . \label{2.33}%
\end{equation}
Together with the Ornstein-Zernicke equation (\ref{2.25}) they are a
closed set of equations to determine the density and pair
correlation function for given potentials. This is the HNC of liquid
state theory, generalized to spatially inhomogeneous systems \cite
{Attard89}. It is known to give very good results for uniform
Coulomb systems \cite {Hansen} and for inhomogenous, confined
Coulomb systems even at strong coupling conditions \cite
{Wrighton09, Wrighton12}.

The corresponding inverse HNC forms for (\ref{2.26}) and
(\ref{2.27})
\begin{equation}
\beta_{c}\mu_{c}(\mathbf{r})=\ln\left(  n\left(  \mathbf{r}\right)
\lambda_{c}^{3}\right)  -\int d\mathbf{r}^{\prime\prime}c^{(2)}(\mathbf{r,r}%
^{\prime\prime}\mid n)n\left(  \mathbf{r}^{\prime\prime}\right)  ,
\label{2.34}%
\end{equation}%
\begin{equation}
\beta_{c}\phi_{c}(\mathbf{r},\mathbf{r}^{\prime})=-\ln(1+h(\mathbf{r,r}%
^{\prime}\mid n))+h(\mathbf{r,r}^{\prime}\mid n)-c(\mathbf{r,r}^{\prime}\mid
n_{c}), \label{2.35}%
\end{equation}
and the Ornstein-Zernicke equation (\ref{2.27}a) is unchanged
\begin{equation}
c^{(2)}\left(  \mathbf{r,r}^{\prime}\mid n\right)  =h\left(  \mathbf{r}%
,\mathbf{r}^{\prime}\mid n\right)  -\int
d\mathbf{r}^{\prime\prime}c^{(2)}\left(
\mathbf{r,r}^{\prime\prime}\mid n\right)  n\left(
\mathbf{r}^{\prime\prime }\right)  h\left(
\mathbf{r}^{\prime\prime}\mathbf{,r}^{\prime}\mid n\right)
. \label{2.36}%
\end{equation}
The "hole function", $h\left(  \mathbf{r},\mathbf{r}^{\prime}\mid n\right)  $
has been introduced for notational simplicity%
\begin{equation}
h\left(  \mathbf{r},\mathbf{r}^{\prime}\mid n\right)  =g\left(  \mathbf{r}%
,\mathbf{r}^{\prime}\mid n\right)  -1. \label{2.36a}%
\end{equation}
These equations, together with (\ref{2.29}) for $\beta_{c}/\beta$, provide
practical forms to determine $\beta_{c}/\beta$, $\beta_{c}\mu_{c}(\mathbf{r}%
)$, and $\beta_{c}\phi_{c}(\mathbf{r},\mathbf{r}^{\prime})$ for the
classical system, given appropriate quantum input.

In the uniform limit (no external potential) the second term of (\ref{2.34})
is simply related to the static structure factor $S(k)$
\begin{equation}
\int d\mathbf{r}^{\prime\prime}c^{(2)}(\mathbf{r,r}^{\prime\prime}\mid
n)n\left(  \mathbf{r}^{\prime\prime}\right)  \rightarrow n\int d\mathbf{r}%
^{\prime\prime}c^{(2)}(\mathbf{r-r}^{\prime\prime},n)=1-\frac{1}{S(k=0)}.
\label{2.36b}%
\end{equation}
For Coulomb systems, such as jellium considered in the following
companion paper \cite {Dutta12} and ideal Fermi fluids at zero temperature,
$S(k)$ vanishes as $k\rightarrow0$
and this contribution to $\beta_{c}\mu_{c}$ diverges. Therefore,
instead of (\ref{2.34}) an alternative form obtained from a coupling
constant integration and the HNC \cite{Baus} is used,
\begin{equation}
\beta_{c}\mu_{c}=\ln\left(  n_{c}\lambda_{c}^{3}\right)  -n\int d\mathbf{r}%
\left(  c(r,n)+\beta_{c}\phi_{c}(r)-\frac{1}{2}h(r,n)\left(
h(r,n)-c(r,n)\right)  \right)  , \label{2.37}%
\end{equation}

\section{Thermodynamics - further considerations}
\label{sec3}
The definition of the equivalent classical system
assures that the pressure and density in the grand ensemble give the
correct quantum results (in principle), e.g.
\begin{equation}
p_{c}(\beta_{c}\mid\mu_{c},\phi_{c})V=-\Omega_{c}(\beta_{c}\mid\mu_{c}%
,\phi_{c})=p(\beta\mid\mu,\phi)V=-\Omega(\beta\mid\mu,\phi) \label{4.1}%
\end{equation}
However, some care is required in the definition of other
thermodynamic properties. For the quantum system the thermodynamic
variables are $\beta ,\mu(\mathbf{r})$ for constant
$\phi(\mathbf{r,r}^{\prime})$. However, their
variation leads to a change in all three classical variables $\beta_{c}%
,\mu_{c}(\mathbf{r}),\phi_{c}(\mathbf{r,r}^{\prime})$, since all are
functions of $\beta$ and functionals of $\mu(\mathbf{r})$.
Therefore, the variation of the pressure leads to (in the following
the volume $V$ is always held constant)
\begin{align*}
\delta\left(  p_{c}V\right)   &  =\left[  \frac{\partial
p_{c}V}{\partial
\beta_{c}}\mid_{\mu_{c},\phi_{c}}+\int d\mathbf{r}d\mathbf{r}^{\prime}%
\frac{\delta p_{c}V}{\delta\phi_{c}(\mathbf{r,r}^{\prime})}\mid_{\beta_{c}%
,\mu_{c}}\frac{\partial\phi_{c}(\mathbf{r,r}^{\prime})}{\partial\beta_{c}}%
\mid_{\mu_{c}}\right]  \delta\beta_{c}\\
&  +\int d\mathbf{r}\left[  \frac{\delta p_{c}V}{\delta\mu_{c}(\mathbf{r}%
)}\mid_{\beta_{c},\phi_{c}}+\int
d\mathbf{r}^{\prime}d\mathbf{r}^{\prime
\prime}\frac{\delta p_{c}V}{\delta\phi_{c}(\mathbf{r}^{\prime}\mathbf{,r}%
^{\prime\prime})}\mid_{\beta_{c},\mu_{c}}\frac{\delta\phi_{c}(\mathbf{r}%
^{\prime}\mathbf{,r}^{\prime\prime})}{\delta\mu_{c}(\mathbf{r})}\mid
_{\beta_{c}}\right]  \delta\mu_{c}(\mathbf{r})
\end{align*}%
\begin{equation}
\equiv\widetilde{S}_{c}dT_{c}+\int d\mathbf{r}\widetilde{n}_{c}(\mathbf{r}%
)\delta\mu_{c}(\mathbf{r}) \label{4.2}%
\end{equation}
The second equality defines the classical thermodynamic entropy and
thermodynamic density in terms of the grand potential
\begin{align}
T_{c}\widetilde{S}_{c}  &  =-\beta_{c}\left[  \frac{\partial p_{c}V}%
{\partial\beta_{c}}\mid_{\mu_{c},\phi_{c}}+\int d\mathbf{r}d\mathbf{r}%
^{\prime}\frac{\delta p_{c}V}{\delta\phi_{c}(\mathbf{r,r}^{\prime})}%
\mid_{\beta_{c},\mu_{c}}\frac{\partial\phi_{c}(\mathbf{r,r}^{\prime}%
)}{\partial\beta_{c}}\mid_{\mu_{c}}\right] \nonumber\\
&
=\beta_{c}\frac{\partial\Omega_{c}}{\partial\beta_{c}}\mid_{\mu_{c}},
\label{4.3}%
\end{align}%
\begin{align}
\hspace{0.25in}\widetilde{n}_{c}(\mathbf{r})  &  =\left[  \frac{\delta p_{c}%
V}{\delta\mu_{c}(\mathbf{r})}\mid_{\beta_{c},\phi_{c}}+\int d\mathbf{r}%
^{\prime}d\mathbf{r}^{\prime\prime}\frac{\delta p_{c}V}{\delta\phi
_{c}(\mathbf{r}^{\prime}\mathbf{,r}^{\prime\prime})}\mid_{\beta_{c},\mu_{c}%
}\frac{\delta\phi_{c}(\mathbf{r}^{\prime}\mathbf{,r}^{\prime\prime})}%
{\delta\mu_{c}(\mathbf{r})}\mid_{\beta_{c}}\right] \nonumber\\
&
=-\frac{\delta\Omega_{c}}{\delta\mu_{c}(\mathbf{r})}\mid_{\beta_{c}},
\label{4.4}%
\end{align}
Similarly, the classical internal energy is defined by%
\begin{align}
\widetilde{E}_{c}  &  \equiv T_{c}\widetilde{S}_{c}+\int d\mathbf{r}%
\widetilde{n}_{c}(\mathbf{r})\mu_{c}(\mathbf{r})-p_{c}V\nonumber\\
&
=\frac{\partial\beta_{c}\Omega}{\partial\beta_{c}}\mid_{\mu_{c}}-\int
d\mathbf{r}\beta_{c}\mu_{c}(\mathbf{r})\frac{\delta\Omega_{c}}{\delta\beta
_{c}\mu_{c}(\mathbf{r})}\mid_{\beta_{c}}\nonumber\\
&  =\frac{\partial\beta_{c}\Omega_{c}}{\partial\beta_{c}}\mid_{\beta_{c}%
\mu_{c}} \label{4.5}%
\end{align}
Note that derivatives in the last equalities of (\ref{4.3}) -
(\ref{4.5}) do not have the restriction of constant $\phi_{c}$.

These same relationships hold for the quantum properties as well,
since they are the general definitions of thermodynamics for the
chosen variables, $\mu,\beta$. In the quantum case $\phi$ is
independent of the thermodynamic variables and hence is constant in
the variations. This leads to the
equivalent expressions in terms of equilibrium averages%
\begin{equation}
n(\mathbf{r})=-\frac{\delta\Omega}{\delta\mu(\mathbf{r})}\mid_{\beta
}=\left\langle \widehat{n}(\mathbf{r})\right\rangle ,\hspace{0.25in}%
E=\frac{\partial\beta\Omega}{\partial\beta}\mid_{\beta\mu}=\left\langle
\widehat{H}\right\rangle .\label{4.6}%
\end{equation}
However, in the classical case the above leads to%
\begin{equation}
\widetilde{n}_{c}(\mathbf{r})=\left\langle \widehat{n}(\mathbf{r}%
)\right\rangle _{c}-\int d\mathbf{r}^{\prime}d\mathbf{r}^{\prime\prime}%
\frac{\delta\Omega_{c}}{\delta\phi_{c}(\mathbf{r}^{\prime}\mathbf{,r}%
^{\prime\prime})}\mid_{\beta_{c},\mu_{c}}\frac{\delta\phi_{c}(\mathbf{r}%
^{\prime}\mathbf{,r}^{\prime\prime})}{\delta\mu_{c}(\mathbf{r})}\mid
_{\beta_{c}}\label{4.7}%
\end{equation}%
\begin{equation}
\widetilde{E}=\left\langle H\right\rangle _{c}+\int
d\mathbf{r}^{\prime
}d\mathbf{r}^{\prime\prime}\frac{\delta\beta_{c}\Omega_{c}}{\delta\phi
_{c}(\mathbf{r}^{\prime}\mathbf{,r}^{\prime\prime})}\mid_{\beta_{c}\mu_{c}%
}\frac{\delta\phi_{c}(\mathbf{r}^{\prime}\mathbf{,r}^{\prime\prime})}%
{\delta\beta_{c}\mu_{c}(\mathbf{r})}\mid_{\beta_{c}}.\label{4.8}%
\end{equation}
For example, the thermodynamic density,
$\widetilde{n}_{c}(\mathbf{r}),$ differs from the average density of
the text above, $n_{c}(\mathbf{r})$, because the latter is defined
as a derivative of the grand potential at constant $\phi_{c}$.

\section{Example - Inhomogeneous Ideal Fermi Gas}

\label{sec4}To illustrate this definition of an equivalent classical
statistical mechanics, consider the case of non-interacting Fermions in an
external potential%
\begin{equation}
H-\mu N\rightarrow K-\int d\mathbf{r}\mu(\mathbf{r})\widehat{n}(\mathbf{r}).
\label{3.1}%
\end{equation}
Since $H-\mu N$ is now the sum of single particle operators, the pressure,
density, and pair correlation function can be expressed in the form of single
particle calculations%
\begin{equation}
p(\beta\mid\mu)V\equiv p(\beta\mid\mu,\phi=0)V=\beta^{-1}(2s+1)\int d\mathbf{r}%
\left\langle \mathbf{r}\right\vert \ln\left(  1+e^{-\beta\left(
\frac{\widehat{p}^{2}}{2m}-\mu(\widehat{\mathbf{r}})\right)
}\right)
\left\vert \mathbf{r}\right\rangle , \label{3.2}%
\end{equation}%
\begin{equation}
n(\mathbf{r,}\beta\mid\mu)\equiv n(\mathbf{r,}\beta\mid\mu,\phi
=0)=(2s+1)\left\langle \mathbf{r}\right\vert \left(  e^{\beta\left(
\frac{\widehat{p}^{2}}{2m}-\mu(\widehat{\mathbf{r}})\right)  }+1\right)
^{-1}\left\vert \mathbf{r}\right\rangle \label{3.3}%
\end{equation}%
\begin{align}
g(\mathbf{r},\mathbf{r}^{\prime}\mathbf{;}\beta &  \mid\mu)\equiv
g(\mathbf{r},\mathbf{r}^{\prime}\mathbf{;}\beta\mid\mu,\phi=0)\nonumber\\
&  =1-\frac{1}{2s+1}\frac{n(\mathbf{r,r}^{\prime})n(\mathbf{r}^{\prime
}\mathbf{,r})}{n(\mathbf{r,r})n(\mathbf{r}^{\prime},\mathbf{r}^{\prime}%
)},\hspace{0.2in}n(\mathbf{r,r}^{\prime})=\left\langle \mathbf{r}\right\vert
\left(  e^{\beta\left(  \frac{\widehat{p}^{2}}{2m}-\mu(\widehat{\mathbf{r}%
})\right)  }+1\right)  ^{-1}\left\vert \mathbf{r}^{\prime}\right\rangle
\label{3.4}%
\end{align}
where $\left\langle \mathbf{r}\right\vert X\left\vert
\mathbf{r}^{\prime }\right\rangle $ denotes a matrix element in
coordinate representation, and $s$ is the spin.

Further reduction of these results to expose the dependence on
$\beta$ and $\mu\left( \mathbf{r}\right) $ is difficult without
solving the eigenvalue
problem for the single particle Hamiltonian $(\widehat{p}^{2}/2m)%
-\mu(\widehat{\mathbf{r}})$. A useful practical approximation that captures
most of the important exchange effects (but not any bound states if supported
by the external potential) is obtained by replacing the operator $\mu
(\widehat{\mathbf{r}})$ by its eigenvalue $\mu(\mathbf{r})$, a "local density
approximation". The expectation values above then can be calculated as simple integrals%

\begin{equation}
p(\beta\mid\mu)\rightarrow\frac{1}{V}\int d\mathbf{r}\left(  2s+1\right)
\frac{1}{h^{3}}\int d\mathbf{p}\frac{\mathbf{p}^{2}}{2m}\left(  e^{\beta
(\frac{p^{2}}{2m}-\mu(\mathbf{r}))}+1\right)  ^{-1}, \label{3.5}%
\end{equation}%
\begin{equation}
n(\mathbf{r,}\beta\mid\mu)\rightarrow\left(  2s+1\right)  h^{-3}\int
d\mathbf{p}\left(  e^{\beta(\frac{p^{2}}{2m}-\mu(\mathbf{r}))}+1\right)
^{-1}, \label{3.5a}%
\end{equation}%
\begin{equation}
n(\mathbf{r,r}^{\prime})=\frac{1}{h^{3}}\int d\mathbf{p}e^{\frac{i}{\hbar
}\mathbf{p\cdot}\left(  \mathbf{r-r}^{\prime}\right)  }\left(  e^{\beta
(\frac{p^{2}}{2m}-\mu(\mathbf{R}))}+1\right)  ^{-1},\hspace{0.25in}%
\mathbf{R}=\frac{\mathbf{r+r}^{\prime}}{2}. \label{3.6}%
\end{equation}
The results (\ref{3.5}) are the familiar finite temperature
Thomas-Fermi approximations for the thermodynamics, while
(\ref{3.6}) is its extension to structure \cite{DT11}. The
expressions for $n(\mathbf{r,}\beta\mid\mu)$ and
$n(\mathbf{r,r}^{\prime})$ are no longer functionals of
$\mu(\mathbf{r})$, but rather local functions of $\mu(\mathbf{r})$
and $\mu(\mathbf{R})$, respectively. The change of variables from
$\beta,\mu$ to $\beta,n$ is
accomplished by inverting (\ref{3.5a}) to obtain $\mu(\mathbf{r}%
)=\mu(\mathbf{r,}\beta\mid n)$. Practical fits for $n(\mathbf{r,}\beta\mid
\mu)$ and this inversion are recalled in Appendix \ref{apB}, along with
simplification of $n(\mathbf{r,r}^{\prime})$.

The corresponding classical results are non-trivial because $\phi
_{c}(\mathbf{q}_{i},\mathbf{q}_{j})\neq0$ when $\phi(\mathbf{q}_{i}%
,\mathbf{q}_{j})=0$, because classical pair interactions are
required to reproduce quantum exchange effects. Thus the
thermodynamics and structure of a simple ideal quantum gas requires
a corresponding classical system with the full complexity of an
interacting many-body system. This classical many-body problem is
addressed here and below in the HNC approximation described above by
(\ref{2.34}) and (\ref{2.35}) together with the Ornstein - Zernicke
equation (\ref{2.36}) and (\ref{2.29}) for $\beta_{c}/\beta$. The
solution for the classical parameters is a straight forward
numerical task, but the results are different for each given
external potential. To simplify the illustration here the following
is further restricted to a uniform ideal Fermi gas
($\mu(\mathbf{r})=\mu$) in the thermodynamic limit. See section 5
below for the non-uniform case of charges in a harmonic trap. For
the uniform gas the equations for the classical system parameters
simplify to
\begin{equation}
\frac{\beta_{c}}{\beta}=\frac{n}{\beta p}\left[  1-\frac{1}{6}n\int
d\mathbf{r}g(r)\mathbf{r}\cdot\nabla\beta_{c}\phi_{c}(r)\right]  . \label{3.7}%
\end{equation}%
\begin{equation}
\beta_{c}\mu_{c}=\ln\left(  n_{c}\lambda_{c}^{3}\right)  -n\int d\mathbf{r}%
\left(  c(r)+\beta_{c}\phi_{c}(r)-\frac{1}{2}h(r)\left(
h(r)-c(r)\right)
\right) , \label{3.8}
\end{equation}
\begin{equation}
\beta_{c}\phi_{c}(r)=-\ln\left(  1+h(r)\right)  +h(r)-c(r), \label{3.9}%
\end{equation}%
\begin{equation}
c\left(  r\right)  =h\left(  r\right)  -n\int d\mathbf{r}^{\prime}c\left(
\left\vert \mathbf{r-r}^{\prime}\right\vert \right)  h\left(  r^{\prime
}\right)  . \label{3.10}%
\end{equation}
The superscript $(2)$ on $c^{(2)}(r)$ and the dependence of $c\left(
r\right)  ,h\left(  r\right)  $ on thermodynamic variables has been
suppressed for simplicity. These last two equations can be solved
for $\beta_{c}\phi _{c}(r)$ using $h(r)$ from (\ref{3.4}) and
(\ref{3.6}) in the uniform limit. With that result,
$\beta_{c}/\beta$ can be calculated from (\ref{3.7}), and then
$\beta_{c}\mu_{c}$ determined from (\ref{2.37}). Further elaboration
is given in Appendix \ref{apB}.

\begin{figure}[h]
\centering
\includegraphics[width=80mm]{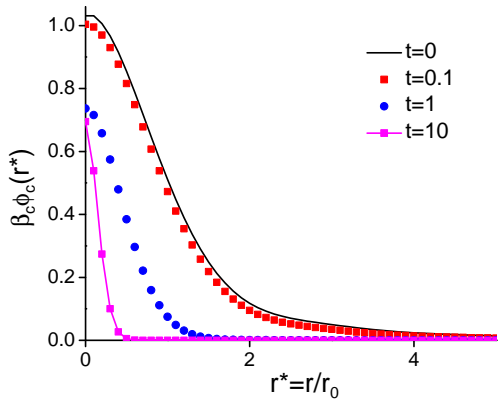}\caption{(color online) Ideal gas Pauli pair potential as a function of $r^* = r/r_0$ for $t = 0, 0.1, 1, 10$.
 }  \label{fig:fig1}
\end{figure}

The dimensionless potential $\beta_{c}\phi_{c}$ will be referred to
as the Pauli potential. It is shown in Figure \ref{fig:fig1} as a
function of the dimensionless coordinate $r^{\ast}=r/r_{0}$ where
$r_{0}$ is the mean distance between particles defined by $4\pi
r_{0}^{3}/3=1/n$. The state conditions are represented by
$r_{s}=r_{0}/a_{B}$ characterizing the density (where $a_{B}$ is the
Bohr radius), and $t=\beta_{F}/\beta$ for the temperature relative
to the Fermi temperature
($\beta_{F}^{-1}=\epsilon_{F}=\hbar^{2}\left(  3\pi ^{2}n\right)
^{2/3}/2m$). For example, in these units $n\lambda^{3}=8/\left(
3\pi^{1/2}t^{3/2}\right)  $ and the classical limit occurs for
$t>>1$ where the distance between particles is large compared to the
thermal de Broglie wavelength. It is noted that ideal gas properties
expressed in terms of $r_{s}$ and $t$ become independent of $r_{s}$
Figure \ref{fig:fig1} shows the Pauli potential at for $t=0$,
$10^{-1}$, $1$, and $10$.

Generally, the potential is positive, finite at $r=0$ (Pauli
exclusion), and monotonically decreasing. The behavior is
exponential at small $r$, but an $r^{-2\text{ }}$algebraic tail
develops for small $t$. This arises from the direct correlation
function $c^{(2)}(r)$ in (\ref{3.9}). Classical statistical
mechanics does not exist for such a long range potential and it
would appear that the equivalent classical system proposed here
fails even for this simplest case of an ideal Fermi gas. However,
this problem is "cured" for the corresponding case of classical
Coulomb interactions with the same long range problem by adding a
uniform neutralizing background (the one component plasma). The same
procedure can be used here, i.e., a classical system is considered
where the pair potential is supplemented in the Hamiltonian by a
corresponding uniform compensating background. The pressure equation
(\ref{2.22}) is modified due to this background by the replacement
of $g_{c}(\mathbf{r,r}^{\prime})$ by
$g_{c}(\mathbf{r,r}^{\prime})-1$, and (\ref{3.7}) becomes
\begin{equation}
\frac{\beta_{c}}{\beta}=\frac{n}{\beta p}\left[  1-\frac{1}{6}n\int
d\mathbf{r}h(r)\mathbf{r}\cdot\nabla\beta_{c}\phi_{c}(r)\right]  .
\label{3.11}%
\end{equation}

\begin{figure}[h]
\centering
\includegraphics[width=80mm]{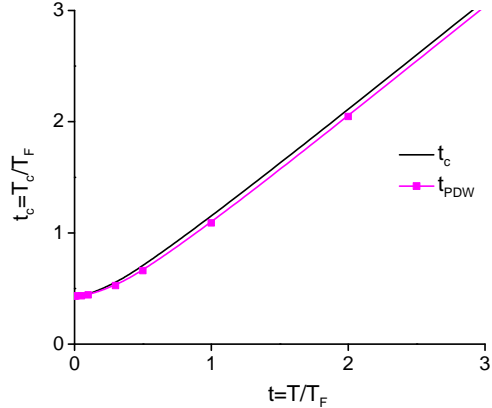}\caption{(color online) Ideal gas reduced classical temperature $t_{c} = T_{c} /T_F$ as a function of $t = T /T_F$ .
Also shown is the result of PDW. } \label{fig2}
\end{figure}

Figure \ref{fig2} shows the classical temperature relative to the
Fermi temperature, $\beta_{F}/\beta_{c}\equiv t_{c}$, as a function
of $t$ obtained from (\ref{3.11}). It is seen that the classical
temperature $T_{c}$ remains finite at $T=0$ in all cases, and
crosses over to $T_{c}=T$ at high temperatures. The PDW model
postulates the form $T_{c}=\left( T^{2}+T_{0}^{2}\right) ^{1/2}$.
The model originally uses the average energy per particle at $T=0$
to evaluate $T_{0}=2T_{F}/5$. The result from (\ref{3.11}) is quite
close $T_{c}\left(  t=0\right) \sim0.43T_{F}.$ To compare the
dependence at finite $t$, the PDW form is also shown in Figure
\ref{fig2} with $T_{0}=T_{c}\left(  t=0\right) $. It is seen that
the results are quite similar although the PDW form has a somewhat
faster cross over to the classical limit.

\begin{figure}[h]
\centering
\includegraphics[width=80mm]{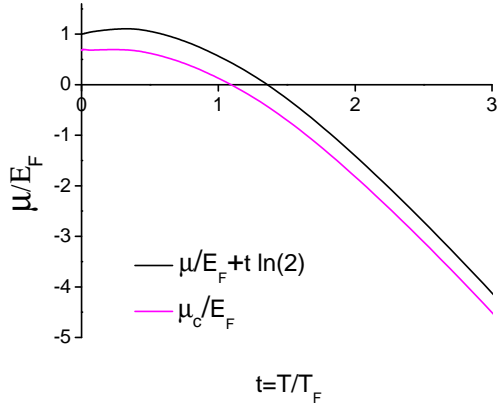}\caption{(color online) Ideal gas dimensionless chemical potential $\mu_c/E_F$ as a function of $t$. Also shown is the corresponding
quantum chemical potential $\mu/E_F + t\ln 2$.} \label{fig3}
\end{figure}

Figure \ref{fig3} shows similar results for $\mu_{c}/E_F$ as a
function of $t$ obtained from (\ref{3.8}). Also shown is the result
for the quantum $\mu/E_F$. Both forms  depend only on $t$
(independent of $r_{s}$). At high temperatures the chemical
potential of the representative system goes over to
$\ln(n\lambda^{3})$ as in equation (\ref{3.8}) while the quantum
chemical potential goes over to $\ln(n\lambda^{3}/2)$. They differ
by a factor of $\ln(2)$ at high temperatures due to the internal
spin degrees of freedom in the quantum calculation. This is
accommodated in the comparison shown in Figure \ref{fig3}.

With the parameters of the equivalent classical system determined, it is now
possible to calculate any desired property of the ideal quantum gas by
classical methods. Since the pressure is given exactly by the definition of
the classical - quantum correspondence equations, it is useful to consider the
internal energy. First, note that all the dimensionless thermodynamic
quantities of both the quantum and classical ideal gas depend on only one
thermodynamic parameter $z=e^{\beta\mu}$. In particular
\begin{equation}
\beta_{c}p_{c}\lambda_{c}^{3}=G(z). \label{3.12}%
\end{equation}
The specific form for $G(z)$ is not required for the present
discussion. The corresponding quantum result is $\beta p
\lambda^{3}=f_{3/2}(z)$, where $f_{3/2}(z)$ is the Fermi integral of
Appendix \ref {apB}, (\ref {b.18}). Although $G(z)$ and $f_{3/2}(z)$
are quite different, the fact that they both depend only on $z$
implies that the relationship among different thermodynamic
properties is the same in both classical and quantum cases. For
example, the energy per particle is determined from the pressure via
the thermodynamic definition (\ref{4.5}) with (\ref{3.8})
\begin{equation}
\widetilde{E}_{c}=-VG(z)\frac{\partial\lambda_{c}^{-3}}{\partial\beta_{c}%
}\mid_{z_{c}}=\frac{2}{3}p_{c}V\label{3.13}%
\end{equation}
The contribution from $\partial
G(z)/\partial\beta_{c}\mid_{z_{c},V}$ vanishes since $z=z(z_{c})$.
This follows on dimensional grounds since $z$ is dimensionless, and
there is no additional energy scale to make $\beta_{c}$
dimensionless. Hence the classical calculation gives the known exact
quantum result. This non-trivial result for a classical interacting
system is a strong confirmation of the effective classical map
defined here. As noted in the last section, it can be verified that
the classical average of the classical Hamiltonian does not give the
correct relationship to the pressure. Instead it is necessary to
define the internal energy thermodynamically, as is done in
(\ref{3.13}). In a similar way it is verified that the relationship
of the classical entropy to the pressure and density is the same as
in the quantum case\begin{equation}
T_{c}S_{c}=\frac{5}{2}p_{c}V-\mu_{c}\overline{N}_{c}. \label{3.14}%
\end{equation}
This is a form of the Sackur-Tetrode equation valid for the quantum
case. It is emphasized that these simple ideal quantum gas results
are being retained for the more complex effective classical system
with pair interactions via the Pauli potential.

\section{Weak Coupling Pair Potential}

\label{sec8}For systems with real interactions $\phi(\mathbf{q}_{1}%
,\mathbf{q}_{2})$ between \ the particles, the classical pair
potential will
have the form%
\begin{equation}
\beta_{c}\phi_{c}(\mathbf{q}_{1},\mathbf{q}_{2})=\left(  \phi_{c}%
(\mathbf{q}_{1},\mathbf{q}_{2})\right)  ^{(0)}+\Delta(\mathbf{q}%
_{1},\mathbf{q}_{2}),\label{8.1}%
\end{equation}
where $\left(  \phi_{c}%
(\mathbf{q}_{1},\mathbf{q}_{2})\right) ^{(0)}$ is the ideal gas
Pauli potential and $\Delta(\mathbf{q}_{1},\mathbf{q}_{2})$ denotes
the contribution to the effective potential from the real pair
potential. Obviously, in the classical limit
$\Delta(\mathbf{q}_{1},\mathbf{q}_{2})\rightarrow\beta\phi(\mathbf{q}%
_{1},\mathbf{q}_{2})$. Another exact limit is the weak coupling
limit for which the direct correlation function becomes proportional
to the potential \cite {Hansen}, or stated inversely,
\begin{equation}
\beta_{c}\phi_{c}(r)\rightarrow-c(r),\hspace{0.2in}\left(
\beta_{c}\phi
_{c}(r)\right)  ^{(0)}\rightarrow-c^{(0)}(r).\label{8.2}%
\end{equation}
Thus a possible approximation incorporating this limit is%
\begin{equation}
\beta_{c}\phi_{c}(r)\rightarrow\left(  \beta_{c}\phi_{c}(r)\right)
^{(0)}-\left(  c(r)-c^{(0)}(r)\right)  ^{(w)},\label{8.3}%
\end{equation}
where $\left(  c(r)-c^{(0)}(r)\right)  ^{(w)}$ denotes a weak
coupling calculation of the direct correlation functions from the
Ornstein - Zernicke equation (\ref{2.25})
\begin{equation}
-\left(  \beta_{c}\phi_{c}(r)\right)  ^{(w)}=h^{(w)}\left(  r\right)
+n\int d\mathbf{r}^{\prime}\left(  \beta_{c}\phi_{c}(\left\vert
\mathbf{r-r}^{\prime
}\right\vert )\right)  ^{(w)}h^{(w)}(r^{\prime}),\label{8.4}%
\end{equation}
with $h^{w}\left(  r\right)  $ being the quantum hole function in
its weak coupling form. In the following companion paper this
approximation is applied to the electron gas (jellium) for which the
weak coupling form is given by the random phase approximation.

To further interpret approximation (\ref{8.3}) consider a uniform
system and rewrite
$\Delta(\mathbf{q}_{1},\mathbf{q}_{2})=\Delta(\left\vert \mathbf{q}%
_{1}-\mathbf{q}_{2}\right\vert )$ in terms of the static structure
factor
\begin{equation}
S\left(  k\right)  =1+n\int d\mathbf{r}e^{i\mathbf{k\cdot r}}h(r),\label{8.5}%
\end{equation}
\begin{align}
\Delta(\left\vert \mathbf{q}_{1}-\mathbf{q}_{2}\right\vert ) &  =\frac{1}%
{n}\int\frac{d\mathbf{k}}{\left(  2\pi\right)
^{3}}e^{-i\mathbf{k\cdot
}\left(  \mathbf{q}_{1}-\mathbf{q}_{2}\right)  }\left(  \frac{1}{S(k)}%
-\frac{1}{S^{(0)}(k)}\right)  ^{(w)}\nonumber\\
&  =\beta\phi(\left\vert \mathbf{q}_{1}-\mathbf{q}_{2}\right\vert
)-\int
d\mathbf{q}G(\mathbf{q})\beta\phi(\left\vert \mathbf{q}_{1}-\mathbf{q}%
_{2}-\mathbf{q}\right\vert )\label{8.6}%
\end{align}
Here $G(\mathbf{q})$ is known as the local field corrections in
linear response for the classical system. It is seen that the
quantum corrections to $\beta\phi(\left\vert
\mathbf{q}_{1}-\mathbf{q}_{2}\right\vert )$ can be interpreted as
local field corrections \cite {STLS,Tanaka86}, although due to quantum
effects rather than classical correlations. In the case of jellium,
the qualitative features of (\ref{8.6}) are a regularization of the
Coulomb singularity at $\mathbf{q}_{1}=\mathbf{q}_{2}$, and
cross-over to an asymptotic Coulomb decay for large $\left\vert
\mathbf{q}_{1}-\mathbf{q}_{2}\right\vert $ with effective coupling
constant given by the exact perfect screening sum rule.

\section{Discussion}
\label{sec7} The objective here has been to define an effective
classical equilibrium statistical mechanics that corresponds to a
chosen quantum system of interest. The motivation is to allow
application of existing strong coupling classical methods (e.g.
liquid state theory, MD simulation) to calculate properties of the
quantum system under conditions for which current theoretical
approaches are not adequate. The correspondence has been defined by
equivalence of the pressures, densities, and pair correlation
functions for the classical and quantum systems. In this way, the
relevant parameters for the classical grand ensemble are fixed - the
temperature, local chemical potential, and pair potential. These
classical parameters were given a more explicit representation in
terms of the quantum parameters by inverting the classical many-body
problem using the HNC integral equations. Formal questions such as
the existence of this inversion have not been addressed, and only
presumed to hold. A counter example is given by the classical
representation of jellium, where the equivalence of pressures is not
possible under conditions of negative pressures.

The three correspondence conditions of section \ref{sec2} are not
unique, and other choices may be preferred in specific applications.
Furthermore, applications require the introduction of appropriate
approximate forms for these correspondence conditions that should be
tailored to the particular system at hand. The special case of a
uniform ideal Fermi gas was illustrated using the HNC integral
equation to determine the parameters of the effective classical
system. A peculiarity is the long range nature of the effective pair
potential at very low temperatures, requiring the introduction of a
compensating uniform background. The resulting classical
thermodynamics was shown to reproduce the exact relationships of
various thermodynamic functionals (e.g., pressure, internal energy,
entropy).

The non-uniform ideal Fermi gas is more interesting and its
representation as a functional of the local density is a fundamental
problem within density functional theory \cite {DFT}. Approximations
such as the Thomas-Fermi representation have limited applicability.
An effective classical representation along the lines described here
would provide access to better approximations, since the functional
dependence on density for the classical system is simple.

A weak coupling approximation for the effective pair potential in
systems with real forces was described in the last section. With
that pair potential known, the effective chemical potential and
effective temperature can be calculated. This is illustrated in the
following companion paper for the uniform electron gas \cite
{Dutta12}. The pair correlation function  is calculated for a wide
range of densities and temperatures, and good agreement is obtained
with diffusion Monte Carlo results at zero temperature and recently
reported Restricted Path Integral Monte Carlo results at finite
temperatures.

A second application in that paper is to shell structure for charges
confined in a harmonic trap. Classically, shell structure arises
only from Coulomb correlations \cite {Wrighton09}. A preliminary
investigation there shows new origins of shell structure due to
diffraction and exchange, even in the absence of Coulomb
correlations (mean field approximation).

\section{Acknowledgements}

This research has been supported by NSF/DOE Partnership in Basic Plasma
Science and Engineering award DE-FG02-07ER54946 and by US DOE Grant DE-SC0002139.

\appendix

\section{Exact Coupled Equations for $n_{c}(\mathbf{r})$ and $g_{c}\left(
\mathbf{r,r}^{\prime}\right)  $.}

\label{apA}

The objective of this appendix is to outline the origin of the exact
equations (\ref{2.23}) - (\ref{2.25}) for the classical density and
pair correlation function. First, make a change of thermodynamic
variables from temperature and chemical potential
$\beta_{c},\mu_{c}(\mathbf{r})$ to temperature and density
$\beta_{c},n_{c}(\mathbf{r})$ by the Legendre transformation%
\begin{equation}
F_{c}(\beta_{c}\mid n_{c})=\Omega_{c}\left(
\beta_{c}\mid\mu_{c}\right)
+\int d\mathbf{r}\mu_{c}\left(  \mathbf{r}\right)  n_{c}\left(  \mathbf{r}%
\right)  ,\label{a.1}%
\end{equation}
where now the free energy is a functional of the classical density
rather than $\mu_{c}\left(  \mathbf{r}\right)  $. Their relationship
is given by the first derivative
\begin{equation}
\frac{\delta F_{c}}{\delta n_{c}(\mathbf{r})}=\mu_{c}(\mathbf{r}).\label{a.2}%
\end{equation}
Here, and throughout this Appendix derivatives are taken at constant
$\phi _{c}$, so the densities involved are those defined as in
(\ref{2.17}). The
free energy is now divided into its ideal gas contribution $F_{c}^{(0)}%
=-\beta_{c}^{-1}\int d\mathbf{r}\left[  1-\ln\left(  n_{c}\left(
\mathbf{r}\right)  \lambda_{c}^{3}\right)  \right]  n_{c}\left(
\mathbf{r}\right)  $, where $\lambda_{c}=\left(  2\pi\hbar^{2}\beta
_{c}/m\right)  ^{1/2}$, and the remainder $F_{c,ex}$ (excess free
energy), so
that (\ref{a.2}) becomes an equation for the density%
\begin{equation}
\ln\left(  n_{c}\left(  \mathbf{r};\beta_{c}\right)
\lambda_{c}^{3}\right)
=\beta_{c}\mu_{c}(\mathbf{r})+c_{c}^{(1)}\left(
\mathbf{r};\beta_{c}\mid
n_{c}\right)  .\label{a.3}%
\end{equation}
Here $c_{c}^{(1)}\left(  \mathbf{r}\mid n_{c}\right)  $ is the first
of a family of functions (direct correlation functions) defined by
derivatives of
the free energy%
\begin{equation}
c_{c}^{(m)}\left(  \mathbf{r}_{1},..,\mathbf{r}_{m};\beta_{c}\mid
n_{c}\right)  \equiv-\beta\frac{\delta^{m}F_{c,ex}}{\delta n_{c}%
(\mathbf{r}_{1})..\delta n_{c}(\mathbf{r}_{m})}\label{a.4}%
\end{equation}

Equation (\ref{a.3}) relates the density $n_{c}(\mathbf{r})$ to a
given
external potential (recall $\mu_{c}(\mathbf{r})=\mu_{c}-\phi_{c,ext,}%
(\mathbf{r})$). Consider now a different external potential given by
$\phi_{c,ext,}(\mathbf{r})+\phi_{c}(\mathbf{r,r}^{\prime})$ and
associated density $n_{c}(\mathbf{r,r}^{\prime})$. This external
potential corresponds to the original one
$\phi_{c,ext,}(\mathbf{r})$ plus a new source of potential of the
same form as would occur if another particle were added at the point
$\mathbf{r}^{\prime}$. It follows that this new density is
proportional to the
pair correlation function for the original system \cite{Hansen}%
\begin{equation}
n_{c}\left(  \mathbf{r,r}^{\prime}\right)  =n_{c}\left(
\mathbf{r}\right)
g_{c}\left(  \mathbf{r,r}^{\prime}\right)  .\label{a.5}%
\end{equation}
The equation corresponding to (\ref{a.3}) for this new external potential is%
\begin{equation}
\ln\left(  n_{c}\left(  \mathbf{r}\right)  g_{c}\left(
\mathbf{r,r}^{\prime
}\right)  \lambda_{c}^{3}\right)  =\beta_{c}\mu_{c}(\mathbf{r})-\beta_{c}%
\phi_{c}(\mathbf{r,r}^{\prime})+c_{c}^{(1)}\left(
\mathbf{r};\beta_{c}\mid
n_{c}g_{c}\right)  .\label{a.6}%
\end{equation}
Finally, subtracting (\ref{a.3}) from (\ref{a.6}) gives the desired
equation for $g_{c}\left(  \mathbf{r,r}^{\prime}\right)  $
\begin{equation}
\ln\left(  g_{c}\left(  \mathbf{r,r}^{\prime}\right)
\lambda_{c}^{3}\right)
=-\beta_{c}\phi_{c}(\mathbf{r,r}^{\prime})+c_{c}^{(1)}\left(  \mathbf{r}%
;\beta_{c}\mid n_{c}g_{c}\right)  -c_{c}^{(1)}\left(
\mathbf{r};\beta_{c}\mid
n_{c}\right)  .\label{a.7}%
\end{equation}
The notation used implies that the functional $c_{c}^{(1)}\left(
\mathbf{r};\beta_{c}\mid\cdot\right)  $ in both (\ref{a.3}) and
(\ref{a.6}) are the same. This follows from density functional
theory where it is demonstrated that the free energy is a universal
functional of the density, the same for all external potentials.
Equations (\ref{2.23}) and (\ref{2.24})
now follow directly from (\ref{a.3}) and (\ref{a.7}) and the identity%
\begin{align}
c_{c}^{(1)}\left(  \mathbf{r};\beta_{c}\mid X\right)   &
=c_{c}^{(1)}\left(
\mathbf{r};\beta_{c}\mid Y\right)  +\int_{0}^{1}d\alpha\partial_{\alpha}%
c_{c}^{(1)}\left(  \mathbf{r};\beta_{c}\mid\alpha X+\left(
1-\alpha\right)
Y\right)  \nonumber\\
&  =c_{c}^{(1)}\left(  \mathbf{r};\beta_{c}\mid Y\right)  +\int_{0}^{1}%
d\alpha\int d\mathbf{r}^{\prime}\frac{\delta c_{c}^{(1)}\left(  \mathbf{r}%
;\beta_{c}\mid\alpha X+\left(  1-\alpha\right)  Y\right)  }{\delta
n_{c}\left(  \mathbf{r}^{\prime}\right)  }\left(  X\left(
\mathbf{r}^{\prime
}\right)  -Y\left(  \mathbf{r}^{\prime}\right)  \right)  \nonumber\\
&  =c_{c}^{(1)}\left(  \mathbf{r};\beta_{c}\mid Y\right)  +\int_{0}^{1}%
d\alpha\int d\mathbf{r}^{\prime}c_{c}^{(2)}\left(  \mathbf{r,r}^{\prime}%
;\beta_{c}\mid\alpha X+\left(  1-\alpha\right)  Y\right)  \left(
X\left( \mathbf{r}^{\prime}\right)  -Y\left(
\mathbf{r}^{\prime}\right)  \right)
\label{a.8}%
\end{align}
with appropriate choices for $X$ and $Y$.

The Ornstein - Zernicke equation (\ref{2.25}) is an identity
obtained as follows. The second functional derivative of the grand
potential is related to
the pair correlation function by%

\begin{equation}
\frac{\delta^{2}\left(  -\beta_{c}\Omega_{c}\right)
}{\delta\mu_{c}\left( \mathbf{r}\right)  \mu_{c}\left(
\mathbf{r}^{\prime}\right)  }=\frac{\delta
n_{c}\left(  \mathbf{r}^{\prime}\right)  }{\delta\mu_{c}\left(  \mathbf{r}%
\right)  }=\beta_{c}n_{c}\left(  \mathbf{r}^{\prime}\right)  \left[
\delta\left(  \mathbf{r-r}^{\prime}\right)  +n_{c}\left(
\mathbf{r}\right) \left(  g_{c}\left(  \mathbf{r,r}^{\prime}\right)
-1\right)  \right]
,\label{a.9}%
\end{equation}
Similarly, the second derivative of the free energy is%
\begin{equation}
\frac{\delta^{2}F_{c}}{\delta n_{c}\left(
\mathbf{r}^{\prime}\right)  \delta n_{c}\left(  \mathbf{r}\right)
}=\frac{\delta\mu_{c}(\mathbf{r})}{\delta n_{c}\left(
\mathbf{r}^{\prime}\right)  }=\beta_{c}^{-1}n_{c}\left(
\mathbf{r}\right)  ^{-1}\left[  \delta\left(
\mathbf{r-r}^{\prime}\right) -n_{c}\left(  \mathbf{r}\right)
c_{c}^{(2)}\left(  \mathbf{r,r}^{\prime}\mid
n_{c}\right)  \right]  .\label{a.10}%
\end{equation}
Then the chain rule
\begin{equation}
\int d\mathbf{r}^{\prime\prime}\frac{\delta n_{c}\left(
\mathbf{r}\right) }{\delta\mu_{c}\left(
\mathbf{r}^{\prime\prime}\right)  }\frac{\delta\mu
_{c}(\mathbf{r}^{\prime\prime})}{\delta n_{c}\left(
\mathbf{r}^{\prime
}\right)  }=\delta\left(  \mathbf{r-r}^{\prime}\right)  \label{a.11}%
\end{equation}
can be written%
\begin{equation}
\int d\mathbf{r}^{\prime\prime}\left[  \delta\left(  \mathbf{r-r}%
^{\prime\prime}\right)  +n_{c}\left(  \mathbf{r}\right)  \left(
g_{c}\left( \mathbf{r,r}^{\prime\prime}\right)  -1\right)  \right]
\left[  \delta\left(
\mathbf{r}^{\prime\prime}\mathbf{-r}^{\prime}\right)  -n_{c}\left(
\mathbf{r}^{\prime\prime}\right)  c_{c}^{(2)}\left(
\mathbf{r}^{\prime\prime
}\mathbf{,r}^{\prime}\mid n_{c}\right)  \right]  =\delta\left(  \mathbf{r-r}%
^{\prime}\right)  .\label{a.12}%
\end{equation}
This gives the Ornstein-Zernicke equation (\ref{2.25}).

\section{Inhomogeneous Ideal Fermi Gas}

\label{apB}

The thermodynamic and structural properties of an inhomogeneous ideal Fermi
gas are straightforward to calculate in a representation that diagonalizes the
effective single particle Hamiltonian%
\begin{equation}
\left(  \frac{\widehat{p}^{2}}{2m}-\mu(\widehat{\mathbf{r}})\right)
\psi_{\mathbf{k}}\left(  \mathbf{r}\right)  =\epsilon_{\mathbf{k}}%
\psi_{\mathbf{k}}\left(  \mathbf{r}\right)  , \label{b.1}%
\end{equation}
where $\mathbf{k}$ labels the corresponding quantum numbers. For Fermions with
spin $s$, the quantum numbers are labeled by $\kappa=\left(  s,\mathbf{k}%
\right)  $. The Hamiltonian in second quantized form is then simply%
\begin{equation}
H=\sum_{\kappa}\epsilon_{\mathbf{k}}a_{\kappa}^{\dagger}a_{\kappa},
\label{b.2}%
\end{equation}
where $a_{\kappa}^{\dagger},a_{\kappa}$ are the creation and annihilation
operators for occupation of the states $\left\{  \psi_{\mathbf{k}}\right\}  $.
Then the pressure is found directly from evaluation of the grand potential
$\Omega_{c}$%
\begin{align}
p(\beta &  \mid\mu)V=\beta^{-1}\sum_{\kappa}\ln\left(  1+e^{-\beta\epsilon_{\mathbf{k}}%
}\right)  =\beta^{-1}Tr^{(1)}\ln\left(  1+e^{-\beta\left(  \frac{\widehat{p}^{2}}%
{2m}-\mu(\widehat{\mathbf{r}})\right)  }\right) \nonumber\\
&  =\left(  2s+1\right)  \beta^{-1}\int d\mathbf{r}\left\langle \mathbf{r}%
\right\vert \ln\left(  1+e^{-\beta\left(  \frac{\widehat{p}^{2}}{2m}%
-\mu(\widehat{\mathbf{r}})\right)  }\right)  \left\vert \mathbf{r}%
\right\rangle , \label{b.3}%
\end{align}
where a coordinate representation has been used in the last expression.

The local density and pair correlation function are obtained from the one an
two particle density matricies. In the diagonal representation these are
\begin{equation}
\rho^{(1)}\left(  \kappa_{1};\kappa_{2}\right)  =\left\langle a_{\kappa_{1}%
}^{\dagger}a_{\kappa_{2}}\right\rangle =\left\langle a_{\kappa_{1}}^{\dagger
}a_{\kappa_{1}}\right\rangle \delta_{\kappa_{1},\kappa_{2}} \label{b.4}%
\end{equation}%
\begin{equation}
\rho^{(2)}\left(  \kappa_{1},\kappa_{2};\kappa_{3},\kappa_{4}\right)
=\left\langle a_{\kappa_{1}}^{\dagger}a_{\kappa_{2}}^{\dagger}a_{\kappa_{4}%
}a_{\kappa_{3}}\right\rangle =\left(  \delta_{\kappa_{1},\kappa_{3}}%
\delta_{\kappa_{2},\kappa_{4}}-\delta_{\kappa_{1},\kappa_{4}}\delta
_{\kappa_{2},\kappa_{3}}\right)  \left\langle a_{\kappa_{1}}^{\dagger
}a_{\kappa_{1}}\right\rangle \left\langle a_{\kappa_{2}}^{\dagger}%
a_{\kappa_{2}}\right\rangle \label{b.5}%
\end{equation}
Where the mean occupation number is
\begin{equation}
\left\langle a_{\kappa}^{\dagger}a_{\kappa^{\prime}}\right\rangle
=\delta_{\kappa,\kappa^{\prime}}\left(  e^{\beta\epsilon_{\mathbf{k}}%
}+1\right)  ^{-1}. \label{b.6}%
\end{equation}
The coordinate representations are%
\begin{align}
\rho^{(1)}\left(  \mathbf{r,}\sigma_{1};\mathbf{r}^{\prime},\sigma_{2}\right)
&  =\sum_{\mathbf{k}_{1},\mathbf{k}_{2}}\psi_{\kappa_{1}}^{\ast}\left(
\mathbf{r}\right)  \psi_{\kappa_{2}}\left(  \mathbf{r}^{\prime}\right)
\left\langle a_{\kappa_{1}}^{\dagger}a_{\kappa_{2}}\right\rangle
=\delta_{\sigma_{1},\sigma_{2}}\left\langle \mathbf{r}\right\vert \ln\left(
e^{\beta\left(  \frac{\widehat{p}^{2}}{2m}-\mu(\widehat{\mathbf{r}})\right)
}+1\right)  ^{-1}\left\vert \mathbf{r}^{\prime}\right\rangle \nonumber\\
&  \equiv\delta_{\sigma_{1},\sigma_{2}}n\left(  \mathbf{r},\mathbf{r}^{\prime
}\right)  \label{b.7}%
\end{align}

\begin{align}
\rho^{(2)}\left(  \mathbf{r}_{1},\sigma_{1},\mathbf{r}_{2},\sigma
_{2};\mathbf{r}_{1}^{\prime},\sigma_{3},\mathbf{r}_{2}^{\prime},\sigma
_{4}\right)   &  =\sum_{\mathbf{k}_{1}..\mathbf{k}_{6}}\psi_{\kappa_{1}}%
^{\ast}\left(  \mathbf{r}_{1}\right)  \psi_{\kappa_{2}}^{\ast}\left(
\mathbf{r}_{2}\right)  \psi_{\kappa_{3}}\left(  \mathbf{r}_{1}^{\prime
}\right)  \psi_{\kappa_{4}}\left(  \mathbf{r}_{2}^{\prime}\right)
\left\langle a_{\kappa_{1}}^{\dagger}a_{\kappa_{2}}^{\dagger}a_{\kappa_{4}%
}a_{\kappa_{3}}\right\rangle \nonumber\\
&  =\delta_{\sigma_{1},\sigma_{3}}\delta_{\sigma_{2},\sigma_{4}}%
\sum_{\mathbf{k}_{1}}\psi_{\kappa_{1}}^{\ast}\left(  \mathbf{r}_{1}\right)
\psi_{\kappa_{1}}\left(  \mathbf{r}_{1}^{\prime}\right)  \left\langle
a_{\kappa_{1}}^{\dagger}a_{\kappa_{1}}\right\rangle \sum_{\mathbf{k}_{2}}%
\psi_{\kappa_{2}}^{\ast}\left(  \mathbf{r}_{2}\right)  \psi_{\kappa_{2}%
}\left(  \mathbf{r}_{2}^{\prime}\right)  \left\langle a_{\kappa_{2}}^{\dagger
}a_{\kappa_{2}}\right\rangle \nonumber\\
&  -\delta_{\sigma_{1},\sigma_{4}}\delta_{\sigma_{2},\sigma_{3}}%
\sum_{\mathbf{k}_{1}}\psi_{\kappa_{1}}^{\ast}\left(  \mathbf{r}_{1}\right)
\psi_{\kappa_{1}}\left(  \mathbf{r}_{2}^{\prime}\right)  \left\langle
a_{\kappa_{1}}^{\dagger}a_{\kappa_{1}}\right\rangle \sum_{\mathbf{k}2}%
\psi_{\kappa_{2}}^{\ast}\left(  \mathbf{r}_{2}\right)  \psi_{\kappa_{2}%
}\left(  \mathbf{r}_{1}^{\prime}\right)  \left\langle a_{\kappa_{2}}^{\dagger
}a_{\kappa_{2}}\right\rangle \label{b.8}%
\end{align}
The diagonal elements are%
\begin{equation}
\rho^{(1)}\left(  \mathbf{r,}\sigma_{1};\mathbf{r},\sigma_{1}\right)
=n\left(  \mathbf{r},\mathbf{r}\right)  \label{b.9}%
\end{equation}%
\begin{align}
\rho^{(2)}\left(  \mathbf{r}_{1},\sigma_{1},\mathbf{r}_{2},\sigma
_{2};\mathbf{r}_{1},\sigma_{1},\mathbf{r}_{2},\sigma_{2}\right)&
=\rho ^{(1)}\left(
\mathbf{r}_{1}\mathbf{,}\sigma_{1};\mathbf{r}_{1},\sigma _{1}\right)
\rho^{(1)}\left(  \mathbf{r}_{2}\mathbf{,}\sigma_{2}
;\mathbf{r}_{2},\sigma_{2}\right)\nonumber\\
&-\delta_{\sigma_{1},\sigma_{2}}\rho ^{(1)}\left(
\mathbf{r}_{1}\mathbf{,}\sigma_{1};\mathbf{r}_{2},\sigma
_{1}\right)\rho^{(1)}\left(  \mathbf{r}_{2}\mathbf{,}\sigma_{2}%
;\mathbf{r}_{1},\sigma_{2}\right)
\end{align} \label{b.9a}
Finally, the density and pair correlation function are identified
from the
summation over spin states%
\begin{equation}
n(\mathbf{r})=\sum_{\sigma_{1}}\rho^{(1)}\left(  \mathbf{r,}\sigma
_{1};\mathbf{r},\sigma_{1}\right)  =\left(  2s+1\right)  n\left(
\mathbf{r},\mathbf{r}\right)  \label{b.10}%
\end{equation}%
\begin{equation}
n(\mathbf{r}_{1})n(\mathbf{r}_{2})g\left(  \mathbf{r}_{1},\mathbf{r}%
_{2}\right)  =\sum_{\sigma_{1},\sigma_{2}}\rho^{(2)}\left(  \mathbf{r}%
_{1},\sigma_{1},\mathbf{r}_{2},\sigma_{2};\mathbf{r}_{1},\sigma_{1}%
,\mathbf{r}_{2},\sigma_{2}\right)  =n(\mathbf{r}_{1})n(\mathbf{r}_{2})-\left(
2s+1\right)  n\left(  \mathbf{r}_{1},\mathbf{r}_{2}\right)  n\left(
\mathbf{r}_{2},\mathbf{r}_{1}\right)  \label{b.11}%
\end{equation}
This gives the results (\ref{3.3}) and (\ref{3.4}).

The local density and pair correlation function are determined from the
function $n\left(  \mathbf{r},\mathbf{r}^{\prime}\right)  $ obtained from the
single particle density matrix (\ref{b.7}),%
\begin{equation}
n(\mathbf{r,r}^{\prime})=\left\langle \mathbf{r}\right\vert \left(
e^{\beta\left(  \frac{\widehat{p}^{2}}{2m}-\mu(\widehat{\mathbf{r}})\right)
}+1\right)  ^{-1}\left\vert \mathbf{r}^{\prime}\right\rangle . \label{b.12}%
\end{equation}
In the local density approximation of the text, $\mu(\widehat{\mathbf{r}%
})\rightarrow\mu(\mathbf{R})$, where $\mathbf{R=}\left(  \mathbf{r+r}^{\prime
}\right)  /2$, this becomes (\ref{3.6})%
\begin{equation}
n(\mathbf{r,r}^{\prime})=\frac{1}{h^{3}}\int d\mathbf{p}e^{\frac{i}{\hbar
}\mathbf{p\cdot}\left(  \mathbf{r-r}^{\prime}\right)  }\left(  e^{\beta
(\frac{p^{2}}{2m}-\mu(\mathbf{R}))}+1\right)  ^{-1}. \label{b.13}%
\end{equation}
Further simplification is possible
to get%
\begin{equation}
n(\mathbf{r,r}^{\prime})\lambda^{3}=\frac{2\lambda}{\pi\left\vert
\mathbf{r-r}^{\prime}\right\vert }\int_{0}^{\infty}dxx\left(  z^{-1}%
(\mathbf{R})e^{x^{2}}+1\right)  ^{-1}\sin\left(  2\sqrt{\pi}x\left\vert
\mathbf{r-r}^{\prime}\right\vert \mathbf{/}\lambda\right)  , \label{b.14}%
\end{equation}
with%
\begin{equation}
z(\mathbf{R})=e^{\beta\mu(\mathbf{R})} \label{b.15}%
\end{equation}
Accordingly the density and pressure simplify to
\begin{equation}
n(\mathbf{r})\lambda^{3}=\left(  2s+1\right)  n(\mathbf{r,r})\lambda
^{3}=\left(  2s+1\right)  f_{3/2}(z(\mathbf{r})), \label{b.16}%
\end{equation}%
\begin{equation}
\beta p\lambda^{3}=\frac{1}{V}\int d\mathbf{r}\left(  2s+1\right)
f_{5/2}(z(\mathbf{r})), \label{b.17}%
\end{equation}
with \ the definitions%
\begin{equation}
f_{3/2}(z)=\frac{4}{\sqrt{\pi}}\int_{0}^{\infty}dxx^{2}\left(  z^{-1}e^{x^{2}%
}+1\right)  ^{-1},\hspace{0.2in}f_{5/2}(z)=\frac{8}{3\sqrt{\pi}}\int
_{0}^{\infty}dxx^{4}\left(  z^{-1}e^{x^{2}}+1\right)  ^{-1}. \label{b.18}%
\end{equation}

\end{document}